\documentclass[11pt]{article}
\usepackage{graphicx}
\usepackage{amsmath}
\usepackage{amsfonts}
\usepackage{amssymb}
\usepackage{epsfig}
\usepackage{times}
\usepackage{cite}
\usepackage{calc}
\usepackage{version}
\usepackage[english]{babel}
\begin{document}

%%%%%%%%%%%%%%%%%%%%%%%%%%%%%%%%%%%%%%%%%%%%%%%%%%%%%%%%%%%%%%%%%%%%%%%%%%%%%%
%%%%%%%%%%%%%%%%%%%%%%%%%%%%%%%%%%%%%%%%%%%%%%%%%%%%%%%%%%%%%%%%%%%%%%%%%%%%%%
\begin{titlepage}

\null

\vskip 1.5cm
%\begin{flushright}
%Report: IFIC/11-18, FTUV-11-0419\\
%\end{flushright}

\vskip 1.cm

{\bf\large\baselineskip 20pt
\begin{center}
\begin{Large}
Three-particle correlations in QCD parton showers
\end{Large}
\end{center}
}
\vskip 1cm

\begin{center}
Redamy P\'erez-Ramos
\footnote{e-mail: redamy.perez@uv.es}, 
Vincent Mathieu
\footnote{e-mail: vincent.mathieu@ific.uv.es} and Miguel-Angel Sanchis-Lozano
\footnote{e-mail: miguel.angel.sanchis@ific.uv.es}\\
\medskip
Departament de F\'{\i}sica Te\`orica and IFIC, 
Universitat de Val\`encia - CSIC\\
Dr. Moliner 50, E-46100 Burjassot, Spain
\end{center}

\baselineskip=15pt

\vskip 3.5cm

{\bf Abstract}: Three-particle correlations in quark and gluon jets
are computed for the first time in perturbative QCD. We give results in the double 
logarithmic approximation and the modified leading logarithmic 
approximation. In both resummation schemes, we use the formalism
of the generating functional and solve the evolution
equations analytically from the steepest descent evaluation of
the one-particle distribution. We thus provide a further test of 
the local parton hadron duality and make predictions for 
the LHC.
\end{titlepage}

\section{Introduction}
The observation of quark and gluon jets has played a crucial role
in establishing QCD as the theory of strong interaction within
the standard model of particle physics. 
The jets, narrowly collimated bundles of hadrons, reflect configurations of quarks 
and gluons at short distances.
Powerful schemes, like the double logarithmic approximation (DLA)
and the modified leading logarithmic approximation (MLLA), which allow for the 
perturbative resummation of soft-collinear and hard-collinear gluons before the hadronization occurs, 
have been developed over the past 30 years (for a review see \cite{Basics}). 
One of the most striking predictions of perturbative QCD,
which follows as a consequence of angular ordering (AO) within the MLLA and
the local parton hadron duality (LPHD) hypothesis \cite{LPHD}, is the existence of the hump-backed 
shape \cite{Basics} of the inclusive energy distribution of hadrons, later confirmed 
by experiments at colliders. Indeed, the shape and normalization of single inclusive distributions
are compared with an experiment; a constant ${\cal K}^{ch}$, which normalizes the number of soft gluons to the
number of charged detected hadrons (mostly pions and kaons), turns out to be close to unity (${\cal K}^{ch}\sim1$), 
giving support to the similarity between parton and hadron spectra \cite{Basics}.
Thus, the study of inclusive observables like the inclusive energy distribution and the
transverse momentum $k_\perp$ spectra of hadrons \cite{PerezRamos:2007cr} has shown that the perturbative 
stage of the process, which evolves from the hard scale or leading parton virtuality $Q\sim E$ to the hadronization
scale $Q_0$, is dominant. In particular, these issues suggest that the hadronization stage 
of the QCD cascade plays a subleading role and, therefore, 
that the LPHD hypothesis is successful while treating one-particle inclusive observables.

The study of particle correlations in intrajet cascades, which are less inclusive observables, 
focuses on providing tests of the partonic dynamics and the LPHD. In \cite{FW},
this observable was computed for the first time at small $x$ 
(energy fraction of the jet virtuality taken away by one parton) in MLLA for particles
staying close to the maximum of the one-particle distribution. In \cite{RPR2}, the
previous solutions were extended, at MLLA, to all possible values of $x$ 
by exactly solving the QCD evolution equations. This observable was measured by the OPAL Collaboration in 
$e^+e^-$ annihilation at the $Z^0$ peak, that is, for $\sqrt{s}=91.2\text{ GeV}$ 
at LEP \cite{Acton:1992gd}. Though the agreement with predictions presented in 
\cite{RPR2} was improved, a discrepancy still subsists pointing out a possible failure of 
the LPHD for less inclusive observables. However, these measurements were redone by the CDF 
Collaboration in $p\bar p$ collisions
at the Tevatron for mixed samples of quark and gluon jets \cite{:2008ec}. 
The agreement with predictions presented in \cite{FW} turned out to be rather good, especially 
for particles having very close energy fractions ($x_1\approx x_2$). A discrepancy between the
OPAL and CDF analysis showed up and still stays unclear. Therefore, the measurement of the 
two-particle correlations at higher energies at the LHC becomes crucial.
Furthermore, going one step beyond, in this article we give predictions for the three-particle 
correlations inside quark and gluon jets. This observable and the two-particle correlations 
can be measured on equal footing at the LHC so as to provide further verifications of the LPHD
for less inclusive observables.

\section{Kinematics and evolution equations}
\begin{figure}
\begin{center}
\includegraphics[height=4.0cm]{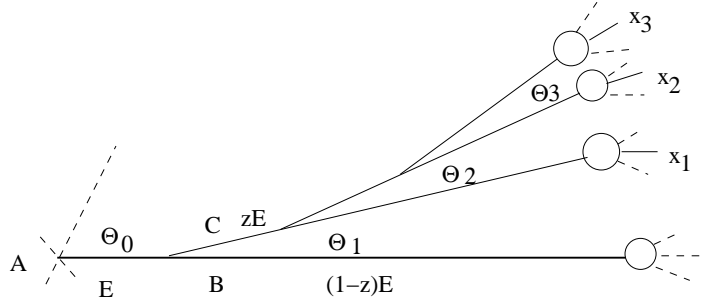}
\caption{Three-particle yield and angular ordering inside a high energy jet.}
\label{fig:three-part}
\end{center}
\end{figure}

A generating functional $Z(E,\Theta;\{u\})$ can be constructed \cite{Basics} 
that describes the azimuth averaged parton content of a jet of energy $E$ with a given opening 
half-angle $\Theta$; by virtue of the exact AO (MLLA), which satisfies an integro-differential
system of evolution equations. In order to obtain {\em exclusive} $n$-particle distributions $D_A^{(n)}(k_i,E)$ 
one takes $n$ variational derivatives of $Z_A$ over $u(k_i)$ with appropriate
particle momenta, $i=1 \ldots n$, and sets $u\equiv 0$ afterwards;
{\em inclusive} distributions are generated by taking variational
derivatives around $u\equiv 1$. Let us introduce the $n$-particle differential
correlations for $A=G,Q,\bar Q$ jets as,
\begin{eqnarray}\label{eq:Anotation}
A_{1\ldots n}^{(n)}(z)\equiv
\frac{x_1}{z}\ldots\frac{x_n}{z}D_A^{(n)}\left(\frac{x_1}{z}\ldots\frac{x_n}{z},\ln\frac{zQ}{Q_0}\right),
\end{eqnarray}
together with $A_{1\ldots n}^{(n)}\equiv A_{1\ldots n}^{(n)}(1)$ for later use; 
$x_i$ corresponds to the Feynman energy fraction of the jet taken 
away by one particle ``$i$" and $z$ is the energy fraction of the intermediate 
parton. For instance, for three-particle correlations $n=3$, 
the observable to be measured reads ${\cal C}^{(3)}_{A_{123}}=\frac{A^{(3)}_{123}}{A_1A_2A_3}$.
The production of three hadrons is displayed in Fig.\ref{fig:three-part} after a 
quark or a gluon ($A$) jet of energy $E$ with half opening angle $\Theta_0$ and virtuality $Q=E\Theta_0$
has been produced in a high energy collision. 
The kinematical variable 
characterizing the process is given by the transverse momentum $k_\perp=zE\Theta_1\geq Q_0$ 
[or $(1-z)E\Theta_1\geq Q_0$] of the first splitting $A\to BC$. The parton $C$ fragments into three offspring such
that three hadrons of energy fractions $x_1$, $x_2$, and $x_3$ can be triggered from the same 
cascade following the condition $\Theta_0\geq\Theta_1\geq\Theta_2\geq\Theta_3$, which arises 
from exact AO in MLLA \cite{Basics}. 
We make use of variables, $\ell=\ln\frac{z}{x_1}$, $y=\ln\frac{x_3E\Theta_1}{Q_0}$, $\ell_i=\ln\frac{1}{x_i}$, 
$y_j=\ln\frac{x_jE\Theta_0}{Q_0}$, $\eta_{ij}=\ln\frac{x_i}{x_j}$, 
$Y=\ell_i+y_j+\eta_{ij}=\ln(Q/Q_0)$ and $\lambda=\ln(Q_0/\Lambda_{QCD})$. 
The two variables entering the 
evolution equations are $z$ and $\Theta_1$, such that $x_1\leq z\leq1\Rightarrow 0\leq\ell\leq\ell_1$.
Accordingly, the anomalous dimension related to the coupling constant can be parametrized as
$$
\gamma_0^2(Q^2)=2N_c\frac{\alpha_s(Q^2)}{\pi},\;\gamma_0^2(\ell+y)
=\frac1{\beta_0(\ell+y+\eta_{ij}+\lambda)},
$$
where $\beta_0=\frac{1}{4N_c}\left(\frac{11}3N_c-\frac43n_fT_R\right)$,
with $T_R=1/2$ and $n_f$ the number of light quark flavors. 
From AO and the initial condition at threshold 
$x_3E\Theta_0\geq x_3E\Theta_1\geq x_3E\Theta_3\geq Q_0$, 
one has the bounds $\frac{Q_0}{x_3E}\leq\Theta_1\leq\Theta_0\Rightarrow0\leq y\leq y_3$
for the integrated evolution equations.
The evolution equations satisfied by (\ref{eq:Anotation}) 
are derived from the MLLA master equation for the generating functional $Z_A(E,\Theta;u(k_i))$. 
For three-particle correlations, one takes the first $\frac{\delta Z_A}{\delta u(k_1)}$, 
second $\frac{\delta^2 Z_A}{\delta u(k_1)\delta u(k_2)}$, and finally third 
$\frac{\delta^3 Z_A}{\delta u(k_1)\ldots\delta u(k_3)}$ 
functional derivatives of $Z_A(E,\Theta;u(k_i))$ over the probing functions 
$u(k_i)$ so as to obtain the differential system of evolution equations:
\begin{eqnarray}\label{eq:mlla3correlqdiff}
\hat{Q}_{\ell y}^{(3)}\!&\!=\!&\!\frac{C_F}{N_c}\gamma_0^2G^{(3)}
-\frac34\frac{C_F}{N_c}\gamma_0^2\left(G^{(3)}_\ell-\beta_0\gamma_0^2G^{(3)}\right),\\
\label{eq:mlla3correlgdiff}
\hat{G}_{\ell y}^{(3)}\!&\!=\!&\!\gamma_0^2G^{(3)}
\!-\!a\gamma_0^2\left(G^{(3)}_\ell\!-\!\beta_0\gamma_0^2G^{(3)}\right)
+(a-b)\gamma_0^2\left[
\left(\hat{G}^{(2)}_{12}G_3\!+\!\hat{G}^{(2)}_{13}G_2+\hat{G}^{(2)_{23}}G_1\right)_\ell\right.\\
\!&\!-\!&\!\left.\beta_0\gamma_0^2\left(\hat{G}^{(2)}_{12}G_3\!+\!\hat{G}^{(2)}_{13}G_2
+\hat{G}^{(2)_{23}}G_1\right)\right]+(a\!-\!c)\gamma_0^2\left[(G_1G_2G_3)_\ell\!-\!
\beta_0\gamma_0^2G_1G_2G_3\right],\notag
\end{eqnarray}
where $\hat{A}^{(2)}_{ij}=A^{(2)}_{ij}-A_iA_j$ and
$\hat{A}^{(3)}=A^{(3)}-A_1A_2A_3-
\hat{A}^{(2)}_{12}A_3-\hat{A}^{(2)}_{13}A_2-\hat{A}^{(2)}_{23}A_1$. The subscripts $\ell$
and $y$ in Eqs. (\ref{eq:mlla3correlqdiff}) and (\ref{eq:mlla3correlgdiff}) 
denote $\partial/\partial\ell$ and $\partial/\partial y$, respectively.
The first terms of Eqs. (\ref{eq:mlla3correlqdiff}) and (\ref{eq:mlla3correlgdiff}) are of 
classical origin and, therefore, universal.
Corrections $\propto-\frac34$, $a$, $(a-b)$, and $(a-c)$, which are ${\cal O}(\sqrt{\alpha_s})$ suppressed, 
better account for energy conservation at each vertex of the splitting process, as compared
with the DLA ${\cal O}(1)$. The hard constants are obtained after integration over the regular part
of the DGLAP splitting functions \cite{Basics} as performed in \cite{FW,RPR2}. In the equation for the gluon 
initiated jet (\ref{eq:mlla3correlgdiff}), the first and second constants $a(n_f=3)=0.935$ and $b(n_f=3)=0.915$ 
were obtained in the frame of the single inclusive 
distribution and two-particle correlations respectively \cite{FW}. 
The third constant $c(n_f)$ appearing for the 
first time in this frame reads
$$
c(n_f)=\frac{1}{4N_c}\!\left[\!\frac{11}3N_c+\frac43n_fT_R\left(1-2\frac{C_F}{N_c}\right)^3\!\right]
\stackrel{n_f=3}{=}0.917.
$$
\subsection{MLLA and DLA solutions of the evolution equations}
Equation (\ref{eq:mlla3correlgdiff}) is self-contained and can be solved iteratively by setting 
$G^{(3)}=C_{G_{123}}^{(3)}G_1G_2G_3$ and $G_{ij}^{(2)}=C_{G_{ij}}^{(2)}G_iG_j$ in the left- and right-hand 
sides of (\ref{eq:mlla3correlgdiff}). Accordingly, the solution of (\ref{eq:mlla3correlqdiff}) 
is also obtained by setting $Q^{(3)}=C_{Q_{123}}^{(3)}Q_1Q_2Q_3$ and $Q_{ij}^{(2)}=C_{Q_{ij}}^{(2)}Q_iQ_j$
in the left-hand side of (\ref{eq:mlla3correlqdiff}) and $G^{(3)}=C_{G_{123}}^{(3)}G_1G_2G_3$ 
in the right-hand side of the same equation such that the iterative solutions 
can be written in the compact form
\begin{eqnarray}\label{eq:sol3partg}
{\cal C}_{A_{123}}^{(3)}=
\left({\cal C}_{A_{12}}^{(2)}\!-\!1\right)\!F_{A_{12}}^{(2)}
\!+\!\left({\cal C}_{A_{13}}^{(2)}\!-\!1\right)\!F_{A_{13}}^{(2)}+
\left({\cal C}_{A_{23}}^{(2)}\!-\!1\right)\!F_{A_{23}}^{(2)}
+\frac{N_c^2}{C_A^2}F_{A_{123}}^{(3)}.
\end{eqnarray}
The MLLA two-particle correlators ${\cal C}_{A_{12}}^{(2)}$ will be taken from \cite{RPR2} 
for the computation of ${\cal C}_{A_{123}}^{(3)}$. Moreover,
\begin{eqnarray}\label{eq:fij}
F_{G_{ij}}^{(2)}\!&\!=\!&\!1+
\frac{1-b\Psi_\ell+\xi_1^{ij}-\epsilon_1}
{2+\Delta_{12}+\Delta_{13}+\Delta_{23}+\epsilon_1},\\
\label{eq:tildef123g}
F_{G_{123}}^{(3)}\!&\!=\!&\!\frac{1-c\Psi_\ell
+\xi_1^{12}+\xi_1^{13}+\xi_1^{23}-\epsilon_1}
{2+\Delta_{12}+\Delta_{13}+\Delta_{23}+\epsilon_1}
\end{eqnarray}
and for the quark jet
\begin{eqnarray}\label{eq:tildefij}
F_{Q_{ij}}^{(2)}\!&\!=\!&\!1+\frac{\tilde\xi_1^{ij}
-\tilde\epsilon_1}{3+\Delta_{12}+\Delta_{13}+\Delta_{23}-a\Psi_\ell+\tilde\epsilon_1},\\
F_{Q_{123}}^{(3)}\!&\!=\!&\!\frac{{\cal C}_{G_{123}}^{(3)}\!\left(1-a\Psi_\ell\right)
+\tilde\xi_1^{12}+\tilde\xi_1^{13}+\tilde\xi_1^{23}
-\tilde\epsilon_1}
{3+\Delta_{12}+\Delta_{13}+\Delta_{23}-\!a\Psi_\ell
+\tilde\epsilon_1},
\label{eq:tildef123}
\end{eqnarray}
where $\Psi_\ell=\psi_{1,\ell}+\psi_{2,\ell}+\psi_{3,\ell}={\cal O}(\gamma_0)$ and 
$\psi=\ln[G(\ell,y)]$. Higher order corrections arising from the solution of the 
system of Eqs. \ref{eq:mlla3correlqdiff} and \ref{eq:mlla3correlgdiff} have been 
neglected in (\ref{eq:sol3partg}). In this case, $G(\ell,y)$ is the inclusive energy distribution, 
which will be inserted from the steepest descent method presented in \cite{RPR2}. The 
other functions appearing in (\ref{eq:fij}) and (\ref{eq:tildef123g}) are 
$\Delta_{ij}=\gamma_0^{-2}\left(\psi_{i,\ell}\psi_{j,y}
+\psi_{i,y}\psi_{j,\ell}\right)={\cal O}(1)$ and
\begin{eqnarray*}
\label{eq:zetaell}
\zeta_\ell\!&\!=\!&\!\frac{\dot{{\cal C}}_{G_{123},\ell}^{(3)}}{\dot{{\cal C}}_{G_{123}}^{(3)}}={\cal O}(\gamma_0^2),\;
\zeta_y=\frac{\dot{{\cal C}}_{G_{123},y}^{(3)}}{\dot{{\cal C}}_{G_{123}}^{(3)}}={\cal O}(\gamma_0^2),\;\\ 
\chi_\ell^{ij}\!&\!=\!&\!\frac{\dot{{\cal C}}_{G_{ij},\ell}^{(2)}}{\dot{{\cal C}}_{G_{ij}}^{(2)}}
={\cal O}(\gamma_0^2),\;
\chi_y^{ij}=\frac{\dot{{\cal C}}_{G_{ij},y}^{(2)}}{\dot{{\cal C}}_{G_{ij}}^{(2)}}={\cal O}(\gamma_0^2),\\
\xi_1^{ij}\!&\!=\!&\!\gamma_0^{-2}\left(\chi_\ell^{ij}\Psi_y+
\chi_y^{ij}\Psi_\ell\right)={\cal O}(\gamma_0),\\
\epsilon_1\!&\!=\!&\!\gamma_0^{-2}\left(\zeta_\ell\Psi_y+
\zeta_y\Psi_\ell\right)={\cal O}(\gamma_0),
\end{eqnarray*}
with $\zeta=\ln \dot{{\cal C}}_{G_{123}}^{(3)}$ and $\chi=\ln \dot{{\cal C}}_{G}^{(2)}$.
The set of functions appearing in (\ref{eq:tildefij}) and (\ref{eq:tildef123}) 
is obtained from the previous by replacing $\zeta\to\tilde\zeta$, $\chi\to\tilde\chi$, $\xi\to\tilde\xi$, 
$\dot{{\cal C}}_{G_{ij}}^{(2)}\to\dot{{\cal C}}_{Q_{ij}}^{(2)}$ and 
$\dot{{\cal C}}_{G_{ij}}^{(3)}\to\dot{{\cal C}}_{Q_{ij}}^{(3)}$ where the dotted $\dot{{\cal C}}_{A_{ij}}^{(2)}$ and
$\dot{{\cal C}}_{A_{ij}}^{(3)}$ are the DLA solutions of the two- and three-particle correlators; that is
why this solution is said to be iterative. Moreover, corrections $\epsilon_1,\tilde\epsilon_1$ and 
$\xi_1^{ij},\tilde\xi_1^{ij}$ are very small and do not play a significant role in the shape and normalization of
the three-particle correlations.

The DLA two-particle correlators are taken from \cite{Dokshitzer:1982ia} and the DLA expression for 
$\dot{{\cal C}}_{A_{ij}}^{(3)}$ can be obtained from (\ref{eq:sol3partg}) by setting all MLLA ${\cal O}(\gamma_0)$
corrections to zero:
\begin{eqnarray}
\label{eq:2partcorrdla}
\dot{{\cal C}}_{A_{ij}}^{(2)}-1\!&\!=\!&\!
\frac{N_c}{C_A}\frac{1}{1+\Delta_{ij}};\\
\label{eq:dla3partsol}
\left(\dot{{\cal C}}_{A_{123}}^{(3)}\!-\!1\right)\!&\!-\!&\!\left(\dot{{\cal C}}_{A_{12}}^{(2)}\!-\!1\right)
\!-\!\left(\dot{{\cal C}}_{A_{13}}^{(2)}\!-\!1\right)\!-\!\left(\dot{{\cal C}}_{A_{23}}^{(2)}\!-\!1\right)\\
\!&\!=\!&\!\frac{N_c}{C_A}\frac{\left(\dot{{\cal C}}_{A_{12}}^{(2)}\!-\!1\right)\!+\!
\left(\dot{{\cal C}}_{A_{13}}^{(2)}\!-\!1\right)
\!+\!\left(\dot{{\cal C}}_{A_{23}}^{(2)}\!-\!1\right)}{2+\Delta_{12}+\Delta_{13}+\Delta_{23}}+
\frac{N_c^2}{C_A^2}\frac1{2+\Delta_{12}+\Delta_{13}+\Delta_{23}}.\notag
\end{eqnarray}
The solutions have the following simple physical interpretation: 
the first term $(=-1)$ in the left-hand side translates the independent or decorrelated 
emission of three hadrons in the shower. After inserting
the two-particle correlator with color factor $\propto\frac{N_c}{C_A}$ (\ref{eq:2partcorrdla}) 
in the left-hand side of (\ref{eq:dla3partsol}),
terms $\propto\frac{N_c}{C_A}$ correspond to the case where two partons 
are correlated inside the same subjet, while the other one is emitted independently from the rest. 
Next, replacing (\ref{eq:2partcorrdla}) in the right-hand side of
(\ref{eq:dla3partsol}), one obtains a contribution $\propto\frac{N_c^2}{C_A^2}$ describing the independent
emission of two partons inside the same subjet. The last term $\propto\frac{N_c^2}{C_A^2}$ 
involves three particles strongly correlated inside the same partonic 
shower as depicted in Fig.\ref{fig:three-part}. This term is indeed the cumulants 
of genuine correlations, first obtained in this article 
for this observable. 

The evaluation of (\ref{eq:sol3partg}), which is expressed in terms of the logarithmic derivatives of
the single inclusive distribution $\ln[G(\ell,y)]$, will be performed using the steepest descent method
to determine $G(\ell,y)$ \cite{Dokshitzer:1982ia,RPR2}. Thus, the MLLA logarithmic derivatives were
written in \cite{RPR2} in the form: 
\begin{eqnarray}\label{eq:psiell}
\psi_{i,\ell}(\mu_i,\nu_i)\!\!&\!\!=\!\!&\!\!\gamma_0e^{\mu_i}+\frac12a\gamma_0^2
\left[e^{\mu_i}\tilde Q(\mu_i,\nu_i)-\tanh\nu_i-\tanh\nu_i\coth\mu_i\Big(1+e^{\mu_i}\tilde Q(\mu_i,\nu_i)\Big)\right]\\
\!\!&\!\!-\!\!&\!\!\frac12\beta_0\gamma_0^2\left[1+\tanh\nu_i\Big(1+K(\mu_i,\nu_i)\Big)+C(\mu_i,\nu_i)
\Big(1+e^{\mu_i}\tilde Q(\mu_i,\nu_i)\Big)\right]+{\cal O}(\gamma_0^2),\cr
\psi_{i,y}(\mu,\nu)\!\!&\!\!=\!\!&\!\!\gamma_0e^{-\mu_i}-\frac12a\gamma_0^2
\left[2+e^{-\mu_i}\tilde Q(\mu_i,\nu_i)+\tanh\nu_i-\tanh\nu_i\coth\mu_i\Big(1+e^{-\mu_i}\tilde Q(\mu_i,\nu_i)\Big)\right]\cr
\!\!&\!\!-\!\!&\!\!\frac12\beta_0\gamma_0^2\left[1+\tanh\nu_i\Big(1+K(\mu_i,\nu_i)\Big)-C(\mu_i,\nu_i)
\Big(1+e^{-\mu_i}\tilde Q(\mu_i,\nu_i)\Big)\right]+{\cal O}(\gamma_0^2),
\label{eq:psiy}
\end{eqnarray}
where the functions $\tilde Q(\mu_i,\nu_i)$, $C(\mu_i,\nu_i)$ and $K(\mu_i,\nu_i)$ 
are defined in \cite{RPR2} and
$(\mu_i,\nu_i)$ are expressed as functions of the original variables $(\ell_i,y_j)$ by inverting 
the nonlinear system of equations \cite{Dokshitzer:1982ia}:
\begin{eqnarray*}
\frac{y_i-\ell_i}{\ell_i+y_i}\!&\!=\!&\!\frac{(\sinh2\mu_i-2\mu_i)-(\sinh2\nu_i-2\nu_i)}{2(\sinh^2\mu_i-\sinh^2\nu_i)},\\
\frac{\sinh\nu_i}{\sqrt{\lambda}}\!&\!=\!&\!\frac{\sinh\mu_i}{\sqrt{\ell_i+y_i+\lambda}}.
\end{eqnarray*} 
In particular, this method allows for the estimation of the observable for particles with
energies near the maximum or hump ($\ell_{max}=Y/2$) of the one-particle distribution 
$\mid\ell-Y/2\mid\ll\sigma\propto Y^{3/2}$, 
which applied to the three-particle correlations will appear in a forthcoming paper. For instance, at DLA
one has $\Delta_{ij}=2\cosh(\mu_i-\mu_j)$ with such a parametrization of the logarithmic derivatives
of the inclusive spectrum. Close to the hump one 
has $\Delta_{ij}\simeq(\ell_i-\ell_j)^2$; thus the correlations are expected to be quadratic as a 
function of $(\ell_i-\ell_j)$ and to have a maximum for particles with the same energy $x_i=x_j$.
In this frame, the role of MLLA corrections should be expected to be larger than for the 
two-particle correlations. Indeed, higher order corrections increase with the
rank of the correlator, which is known from the Koba-Nielsen-Olesen 
problem for intrajet multiplicity fluctuations 
\cite{DokKNO}. For the two-particle correlations, for instance, one has 
$\propto-b(\psi_{1,\ell}+\psi_{2,\ell})$ and for the three-particle correlator one 
has the larger correction $\propto-c(\psi_{1,\ell}+\psi_{2,\ell}+\psi_{3,\ell})$.

\subsection{Phenomenology and comparison with existing $\boldsymbol{e^+e^-}$ and 
$\boldsymbol{p\bar{p}}$ data}
The study of $n$-particle correlations is very important because, being defined as the 
$n$-particle cross section normalized by the product of the single inclusive 
distribution of each parton 
$$
{\cal C}^{(n)}_{A_{1\ldots n}}=\frac{A^{(n)}_{1\ldots n}}{A_1\ldots A_n},
$$ 
the resulting observable becomes 
independent of the constant ${\cal K}^{ch}$, thus providing a refined test
of QCD dynamics at the parton level. Since our study of three-particle correlations
depends on previous results for two-particle correlations, we briefly review recent
results about this observable. The MLLA evolution equations for two-particle correlations, 
quite similar to those leading to the hump-backed plateau, were solved
iteratively in terms of the logarithmic derivatives of $G(\ell,y)$ \cite{RPR2}. 
That is how, the result previously
obtained by Fong and Webber in \cite{FW}, only valid in the vicinity
of the maximum $\ell_{max}$ of the distribution, was extended to all possible values
of $x$. 
\begin{figure}[h]
\begin{center}
\includegraphics[height=5.5cm,width=7.5cm]{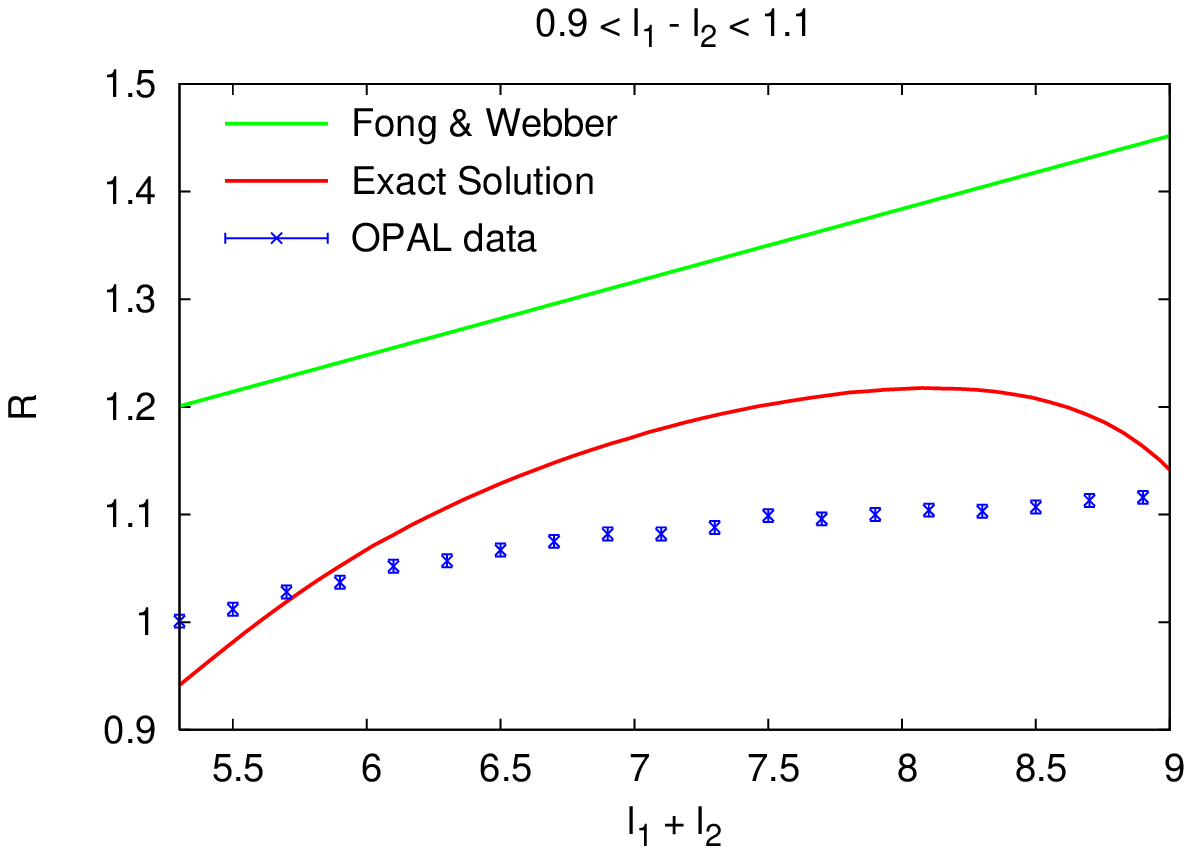}
\includegraphics[height=5.5cm,width=7.5cm]{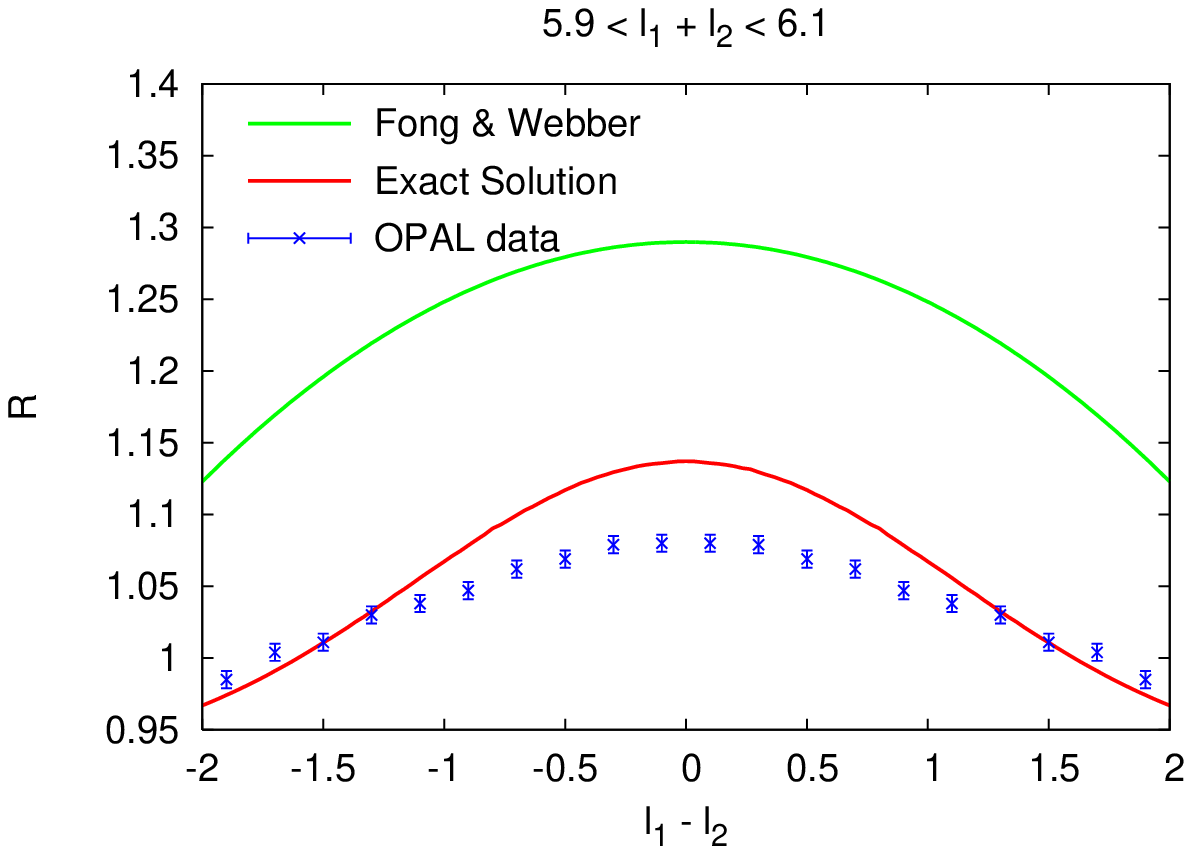}
\caption{\label{fig:opalcorr} Two-particle correlations in two quark 
jets $\left(R=\frac12+\frac12{\cal C}^{(2)}_Q\right)$ \cite{RPR2} in the process $e^+e^-\to q\bar{q}$ as a function of 
$\ell_1+\ell_2=|\ln(x_1x_2)|$ for $\ell_1-\ell_2=\ln(x_2/x_1)=1.0$ (left)
and $\ell_1-\ell_2=|\ln(x_2/x_1)|$ for $\ell_1+\ell_2=\ln(x_1x_2)=6.0$ (right).}  
\end{center}
\end{figure}
Consequently, as displayed in Fig.\ref{fig:opalcorr}, the normalization of the more accurate solution of the
evolution equations is lower and reproduces some features of the OPAL data at the $Z^0$ peak $Q=91.2$ GeV
of the $e^+e^-$ annihilation, like the flattening of the slopes towards smaller values of $x$ \cite{RPR2}. 
Qualitatively, our MLLA expectations agree better
with available OPAL data than the Fong--Webber predictions \cite{RPR2}. 
There remains however a significant discrepancy, markedly at very small $x$. In this region
nonperturbative effects are likely to be more pronounced. They may
undermine the applicability to particle correlations of the LPHD 
considerations that were successful in translating parton level predictions 
to hadronic observations in the case of more inclusive single particle energy 
spectra \cite{Basics}. 

\begin{figure}[h]
\begin{center}
\includegraphics[height=5.5cm,width=7.5cm]{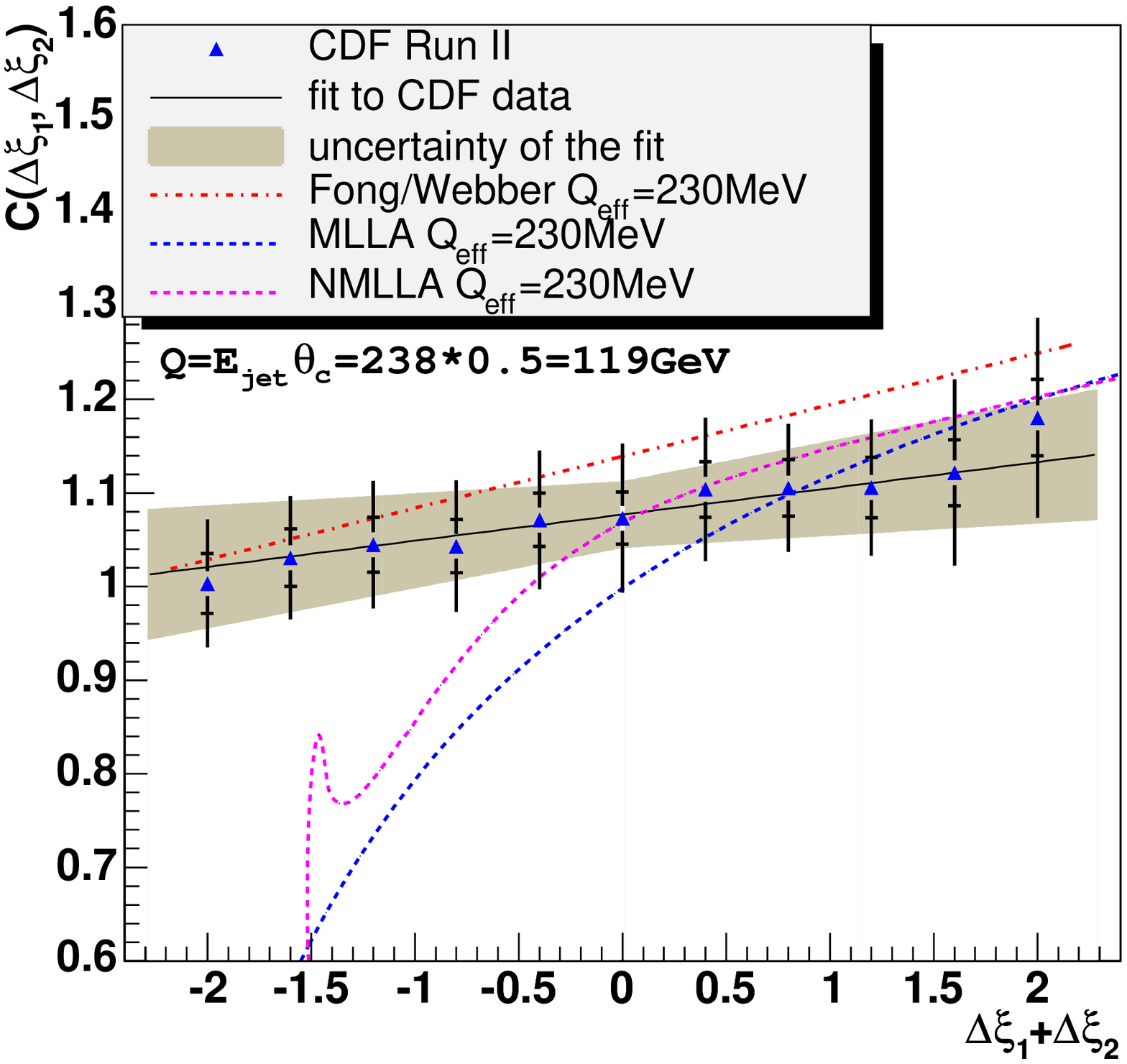}
\includegraphics[height=5.5cm,width=7.5cm]{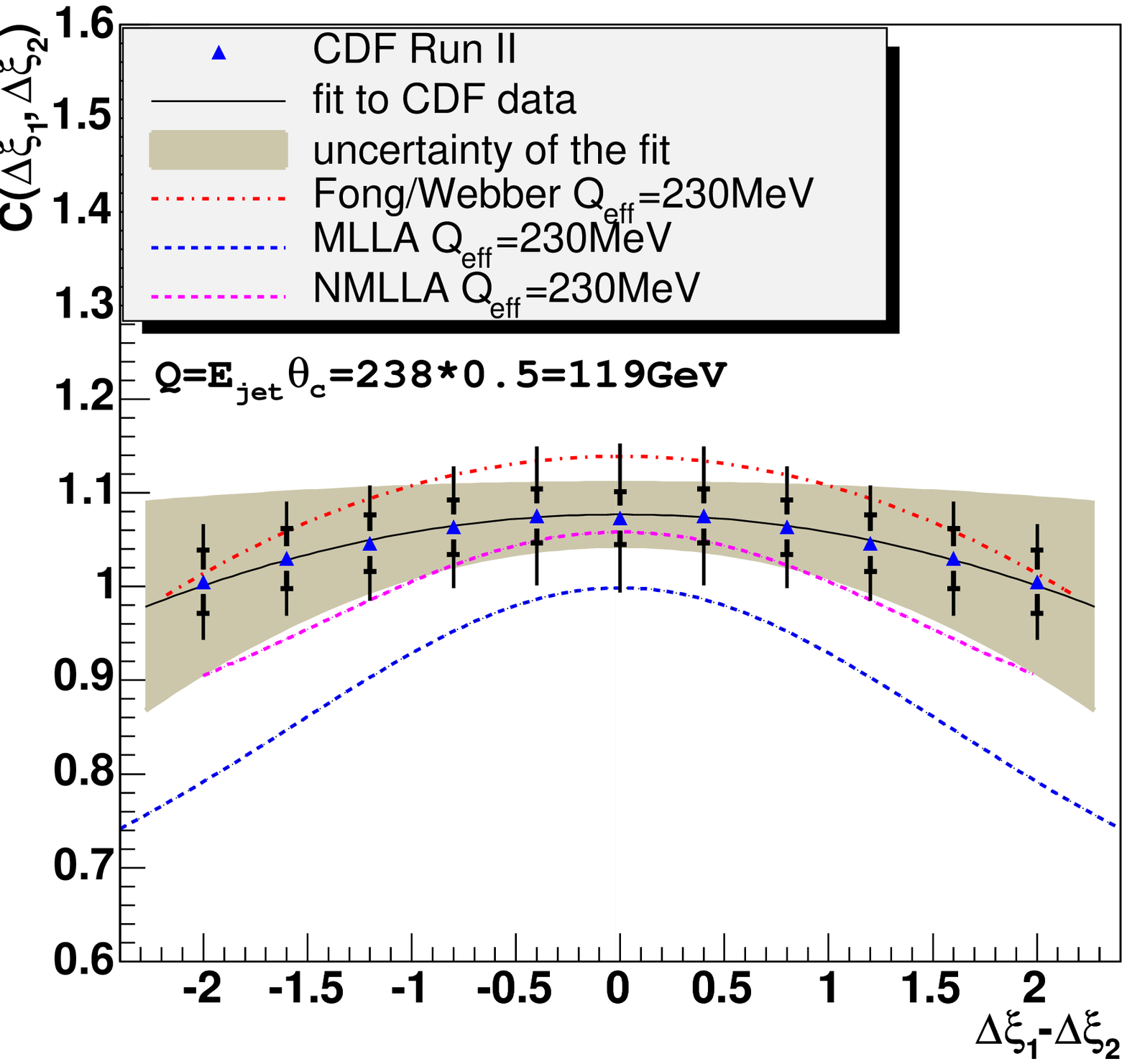}
\caption{\label{fig:cdfcorr} Two-particle correlations in a mixed sample
of gluon and quark jets in $p\bar{p}$ collisions as a function of 
$\Delta\xi_1+\Delta\xi_2=|\ln(x_1x_2)|-\ln(Q/Q_0)$ for $\Delta\xi_1=\Delta\xi_2$ (left)
and $\Delta\xi_1-\Delta\xi_2=|\ln(x_2/x_1)|$ for $\Delta\xi_1=-\Delta\xi_2$ (right).}  
\end{center}
\end{figure}

These measurements were redone by the CDF Collaboration for $p\bar{p}$ collisions at
center of mass energy $\sqrt{s}=1.96$ TeV for mixed samples of quark and gluon jets \cite{:2008ec}.
For comparison with CDF data, the two-particle correlator was normalized by the corresponding multiplicity 
correlator of the second rank, which defines the dispersion of the mean average 
multiplicity inside the jet. In this case, the MLLA solution by Fong and Webber \cite{FW}, 
the more accurate MLLA solution \cite{RPR2}, and the 
NMLLA solution \cite{PerezRamos:2007cr} were compared with the CDF data. The Fong-Webber predictions 
turned out to be in good agreement with CDF data in a range from large to small $x$, also covering 
the region of the phase space where MLLA 
predictions should normally not be reliable, that is, for $x>0.1$ (see Fig.\ref{fig:cdfcorr}).
As these figures were taken from \cite{:2008ec}, different notations have been used in this case, 
for instance, $\ell\equiv\xi=\ln(1/x)$, $\Delta\xi=\xi-\xi_{max}$ 
($\xi_{max}\equiv\ell_{max}=\frac12\ln(Q/Q_0)$) such 
that $\Delta\xi_1+\Delta\xi_2=\ell_1+\ell_2-\ln(Q/Q_0)$ and $\Delta\xi_1-\Delta\xi_2=\ell_1-\ell_2$.

As observed in Fig.\ref{fig:cdfcorr} (left), 
the data are well described by the three cases in the 
interval $\Delta\xi_1+\Delta\xi_2>-0.5$, that is, at very small $x$. However, the Fong and Webber's 
solution also describes the data for $\Delta\xi_1+\Delta\xi_2<-0.5$, that is, for larger values of 
$x$ where the MLLA is no longer valid. QCD color coherence for Fig.\ref{fig:cdfcorr} (left,
the peak at $\Delta\xi_1+\Delta\xi_2=-1.5$ is due to numerical uncertainties) should
be observed if the analysis is extended to $\Delta\xi_1+\Delta\xi_2>2.5$.
Moreover, the NMLLA solution \cite{PerezRamos:2007cr} extends, like for the 
$k_\perp$ spectra, the region of applicability of such predictions for larger values of $x$.
In \cite{:2008ec}, it was concluded that despite the disagreement with the OPAL data in
Fig.\ref{fig:opalcorr}, the LPHD stays successful for the description of less inclusive 
energy-momentum correlations. Therefore, in this paper we encourage the analysis of these
observables by other collaborations like ALICE, ATLAS, and CMS in order to clarify this
mismatch. 
\begin{figure}[h]
\begin{center}
\includegraphics[height=5.5cm,width=7.5cm]{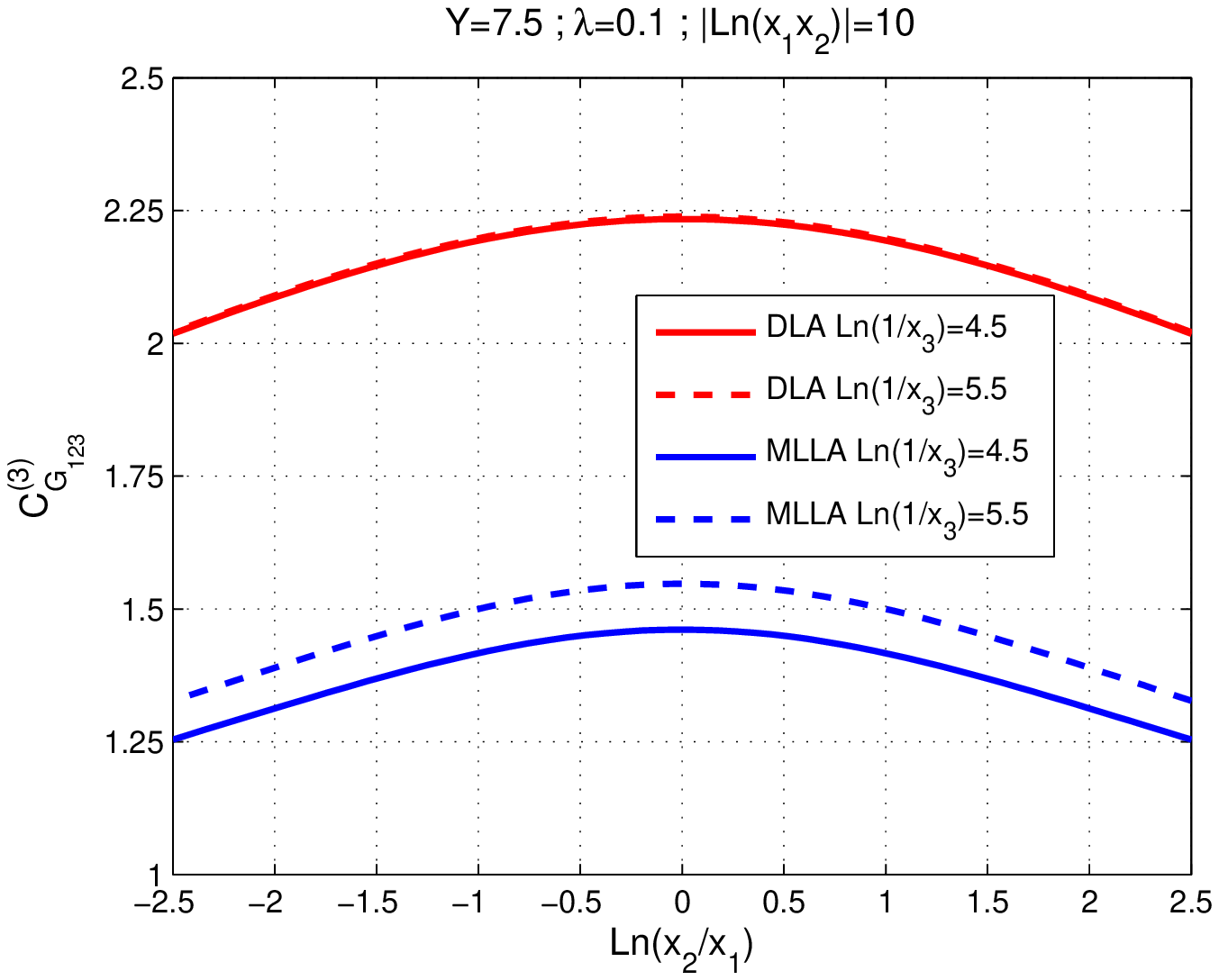}
\includegraphics[height=5.5cm,width=7.5cm]{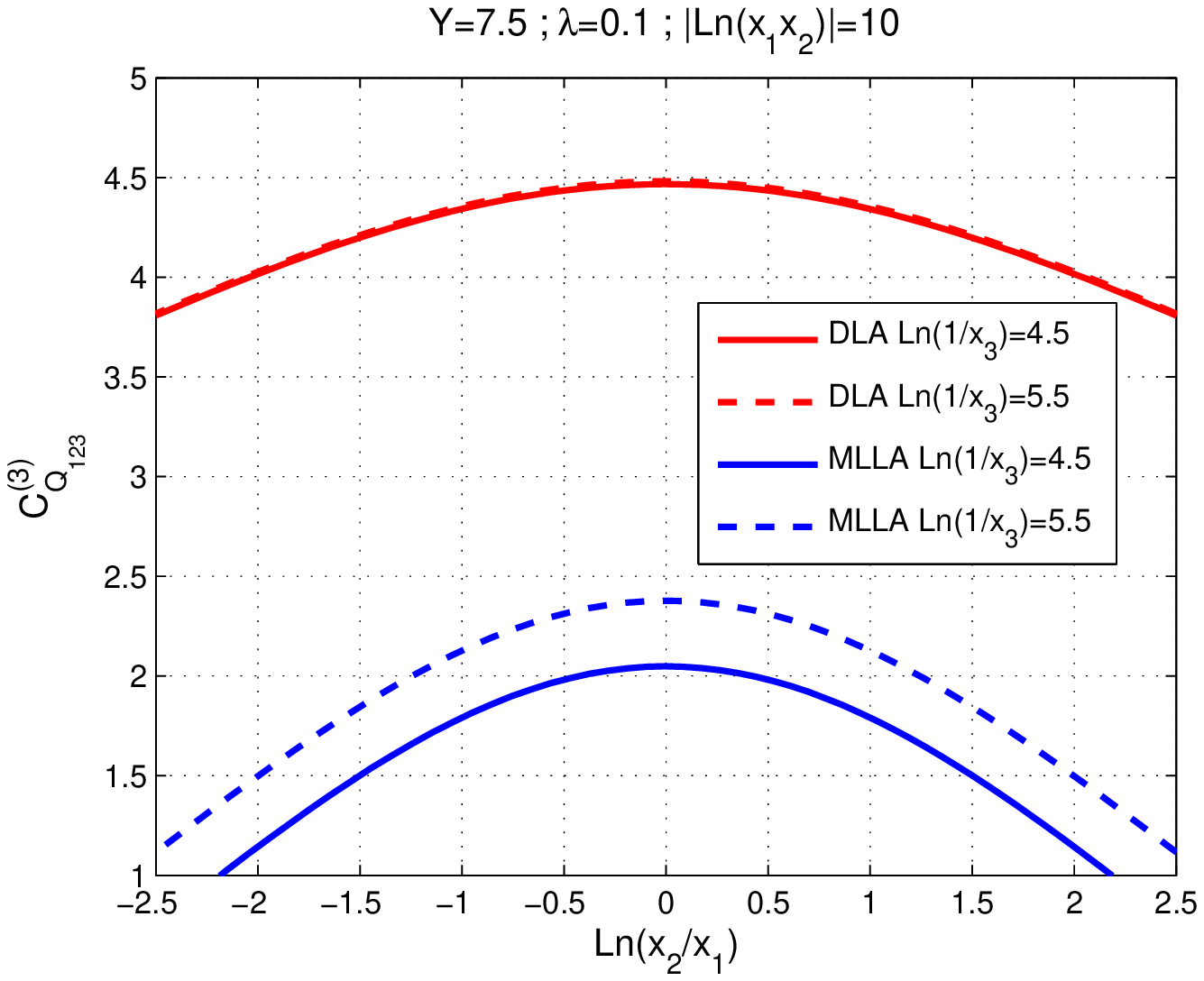}
\caption{\label{fig:3partcorr} Gluon jet 3-particle correlator as a function of 
$|\ln(x_1x_2)|$ for $x_1=x_2$ 
and $\ln(1/x_3)$ (left) and as a function of
$\ln(x_2/x_1)$ for fixed $|\ln(x_1x_2)|$ 
and $\ln(1/x_3)$ (right).}  
\end{center}
\end{figure}
\section{Predictions for three-particle correlations and phenomenology}
Finally, in order to extend the applicability of the LPHD to a larger domain
of observables, we perform theoretical predictions for three-particle correlations 
in the limiting spectrum approximation ($Q_0\approx\Lambda_{QCD}$). This observable and
two-particle correlations can be measured on equal footing at the LHC.
We display the MLLA solutions (\ref{eq:sol3partg})
of the evolution equations (\ref{eq:mlla3correlqdiff}) and (\ref{eq:mlla3correlgdiff}). 
The correlators are functions of 
the variables $\ell_i$, $y_i$ and the virtuality of the jet $Q=E\Theta_0$. After setting 
$y_i=Y-\ell_i$ with fixed $Y=\ln(Q/Q_0)$ in the arguments of the solutions 
(\ref{eq:sol3partg}), the dependence can be reduced to the following: ${\cal C}^{(3)}_{G_{123}}(\ell_1,\ell_2,\ell_3,Y)$ 
and ${\cal C}^{(3)}_{Q_{123}}(\ell_1,\ell_2,\ell_3,Y)$.

In Fig. \ref{fig:3partcorr}, the DLA (\ref{eq:dla3partsol}) and 
MLLA (\ref{eq:sol3partg}) three-particle correlators for $A=G$ and $A=Q,\bar Q$,
$$
{\cal C}^{(3)}_{G_{123}}=\frac{G^{(3)}_{123}}{G_1G_2G_3},\quad
{\cal C}^{(3)}_{Q_{123}}=\frac{Q^{(3)}_{123}}{Q_1Q_2Q_3} 
$$ 
are displayed, respectively, as a function of the difference $(\ell_1-\ell_2)=\ln(x_2/x_1)$
for two fixed values of $\ell_3=\ln(1/x_3)=4.5,\,5.5$, fixed sum $(\ell_1+\ell_2)=|\ln(x_1x_2)|=10$, 
and, finally, fixed $Y=7.5$ (virtuality $Q=450$ GeV and $\Lambda_{QCD}=250$ MeV), which is realistic for LHC 
phenomenology \cite{RPR2}. The representative values $\ell_3=\ln(1/x_3)=4.5,\,5.5$ ($x_3=0.011,\,x_3=0.004$) 
have been chosen according to the range of the energy fraction $x_i\ll0.1$, 
where the MLLA scheme can only be applied. 

\begin{figure}[h]
\begin{center}
\includegraphics[height=5.5cm,width=7.5cm]{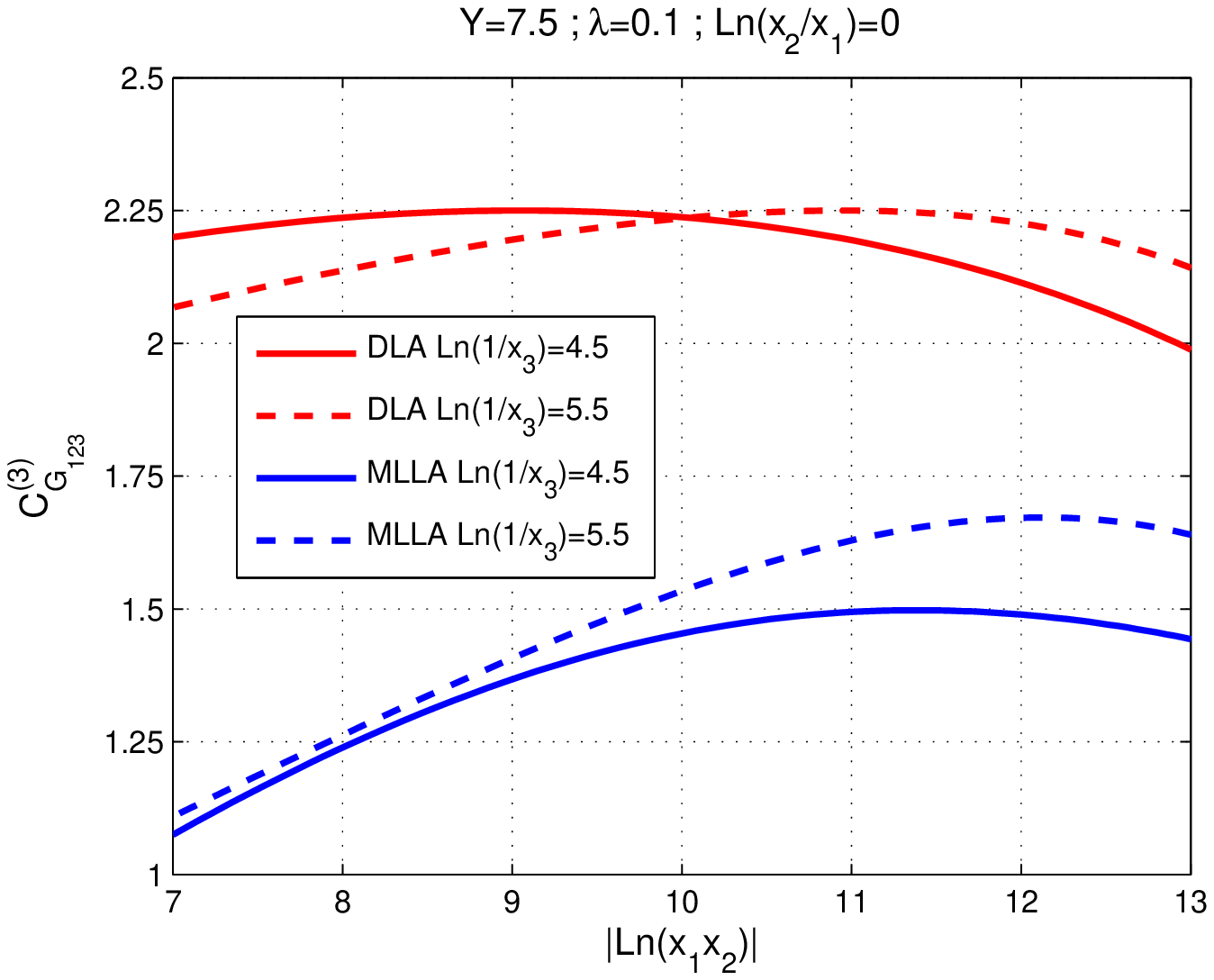}
\includegraphics[height=5.5cm,width=7.5cm]{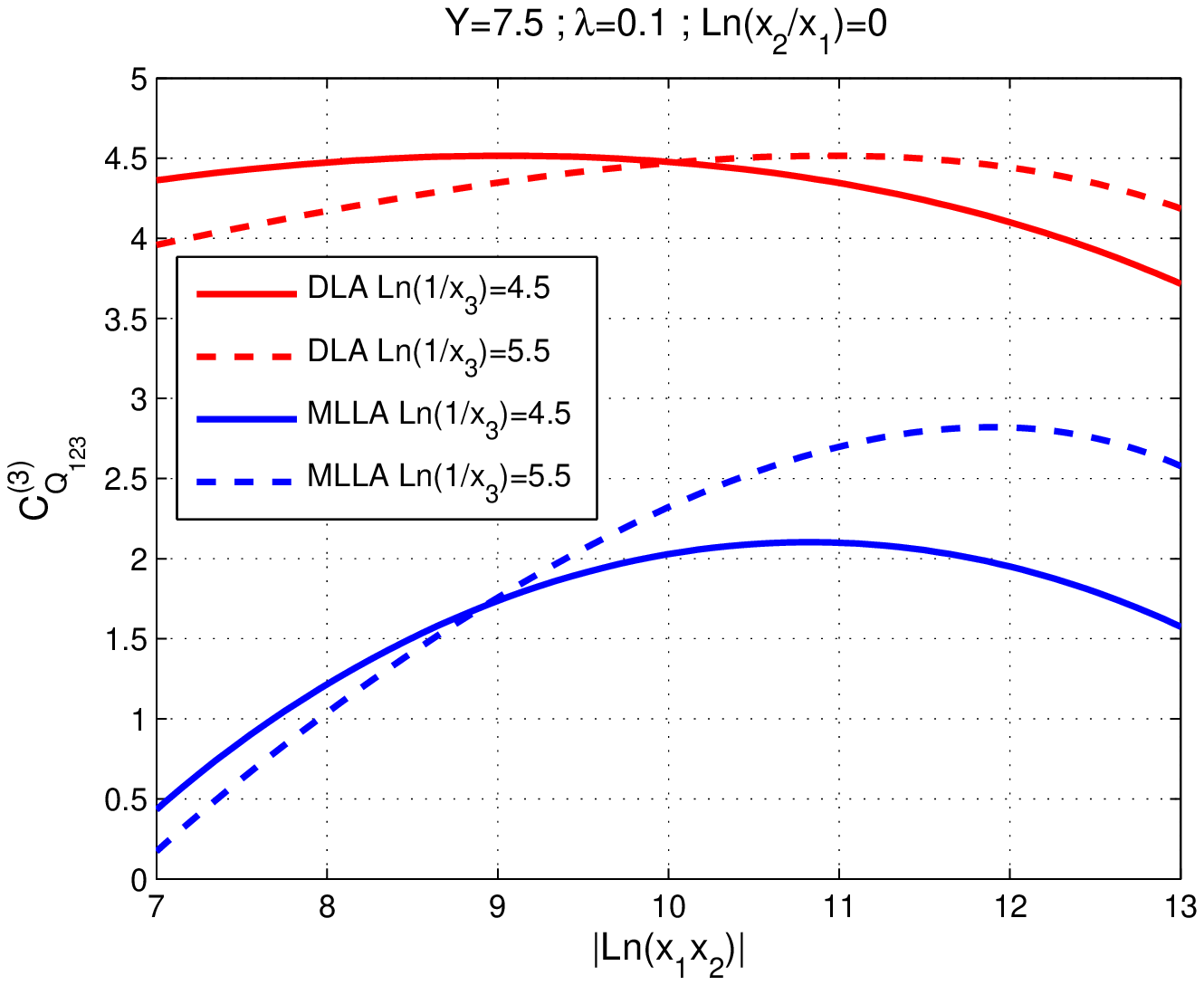}
\caption{\label{fig:3partcorrsom} Quark jet 3-particle correlator as a function of 
$|\ln(x_1x_2)|$ for $x_1=x_2$ 
and $\ln(1/x_3)$ (left) and as a function of
$\ln(x_2/x_1)$ for fixed $|\ln(x_1x_2)|$ 
and $\ln(1/x_3)$ (right).}  
\end{center}
\end{figure}

In Fig. \ref{fig:3partcorrsom}, 
the DLA (\ref{eq:dla3partsol}) and MLLA (\ref{eq:sol3partg}) three-particle correlators 
for $A=G$ and $A=Q,\bar Q$ are depicted, in this case as a function of the sum 
$(\ell_1+\ell_2)=|\ln(x_1x_2)|$ for the same values of $\ell_3=\ln(1/x_3)=4.5,\,5.5$, for $x_1=x_2$ and 
$Y=7.5$. As expected in both cases, the DLA and MLLA three-particle correlators are larger inside a 
quark than in a gluon jet. Of course, these plots will be the same and the interpretation will
apply to all possible permutations of three particles (123). As remarked above, 
the difference between the DLA and MLLA results is quite important in pointing out 
that overall corrections in ${\cal O}(\sqrt{\alpha_s})$ are large. Indeed, the last 
behavior is not surprising as it was already observed in the treatment of 
multiplicity fluctuations of the third kind given by 
$R_3=4.52\left[1-(2.280-0.018n_f)\sqrt{\alpha_s}\right]$ \cite{MALAZA}.

For instance, for one quark jet produced at the $Z^0$ peak of the $e^+e^-$ annihilation 
($Q=45.6$ GeV), one has $\alpha_s=0.134$. Replacing this value into the previous 
formula for the quark jet multiplicity correlator, one obtains a variation 
from 4.52 (DLA) to 0.83 (MLLA). Because of this,
DLA has been known to provide unreliable predictions which should not be compared
with experiments. From Fig.\ref{fig:3partcorr}, the 
correlation is observed to be the strongest when particles have the same energy and 
to decrease when one parton is harder than the others. Indeed, in this region of the phase 
space two competing constraints should be satisfied: as a consequence of gluon coherence and AO, 
gluon emission angles should decrease and on the other hand, the convergence of the perturbative series 
$k_\perp=x_iE\Theta_i\geq Q_0$ should be guaranteed. That is why, as the collinear cutoff 
parameter $Q_0$ is reached, gluons are emitted at larger angles and destructive interferences 
with previous emissions occur. This effect is clearly observed in Fig. \ref{fig:3partcorr}; the steep fall of
the distribution is more pronounced in the quark jet than in the gluon jet.
Moreover, the observable increases for softer partons with 
$x_3$ decreasing, which is for partons less sensitive to the energy balance. In 
Fig.\ref{fig:3partcorrsom} the MLLA correlations increase 
for softer partons, then flatten and decrease as a consequence of soft gluon coherence, 
reproducing for three-particle correlations the hump-backed shape of the one-particle
distribution. Because of the limitation of phase space, one has ${\cal C}^{(3)}\leq1$ for 
harder partons. 

\section{Summary}
In this paper we provide the first full perturbative QCD treatment of three-particle
correlations in parton showers, provide a further test of the LPHD within the
limiting spectrum approximation, and briefly revise the comparison of two-particle
correlations with OPAL and CDF data. The correlations have been shown to
be strongest for the softest hadrons having the same energy $x_1=x_2=x_3$ in both quark and gluon jets,
increasing as a function of $\ln(x_i/x_j)$ and $|\ln(x_ix_j)|$ when $x_k$ softens, that is for partons
less sensitive to the energy balance. This result becomes therefore universal for $n$-particle
correlations.

Coherence effects appear when one or two of the partons involved in the process 
are harder than the others, thus reproducing for this observable the hump-backed shape of the
one-particle distribution. Also, the two- and three-particle correlations vanish 
(${\cal C}^{(2)}\to1$) when one of the partons becomes very soft, thus describing the 
hump-backed shape of the one-particle distribution. The reason
for that is dynamical rather than kinematical: radiation of a soft
gluon occurs at large angles, which makes the radiation
coherent and thus insensitive to the internal parton structure of the
jet ensemble.

We give the first analytical predictions of
this observable in view of forthcoming measurements by ATLAS, CMS, and ALICE at the LHC. 
Further information from the comparison with forthcoming data may also help to improve
Monte Carlo event generators in the soft region of the phase space in intrajet
cascades, where PYTHIA, ARIADNE and HERWIG face difficulties while reproducing the 
data \cite{Abbiendi:2006qr}.

\vskip 1.5cm

We gratefully acknowledge enlightening discussions with W. Ochs and 
E. Sarkisyan-Grinbaum as well as 
support from Generalitat Valenciana under Grant No. PROMETEO/2008/004.
M.A.S. acknowledges support from FPA2008-02878 and GVPROMETEO2010-056. V.M acklowledges support from
the grant HadronPhysics2, a FP7-Integrating 
Activities and Infrastructure Program of the
European Commission under Grant No. 227431, by UE (Feder), and by the MICINN 
(Spain) Grant No. FPA 2010-21750-C02-01.

%%%%%%%%%%%%%%%%%%%%%%%%%%%%%%%%%%%%%%%%%%%%%%%%%%%%%%%%%%%%%%%%%%%%%%%%%%%
%
\end{document}